\documentclass[a4paper]{jpconf}

\usepackage{graphicx}

\begin{document}
\title{The Status of the MicroBooNE Experiment}

\author{B J P Jones}

\address{Laboratory for Nuclear Science, Massachusetts Institute of Technology, 77 Massachusetts Ave., Cambridge, MA 02139, United States of America}

\ead{bjpjones@mit.edu}

\begin{abstract}
The MicroBooNE experiment is an upcoming liquid argon TPC experiment which will pursue a program of short baseline neutrino physics in the Fermilab Booster Neutrino Beam, starting in early 2014.  I give a description of the experiment and review the current status of various aspects of the MicroBooNE detector.
\newline\newline
\textit{contribution to NUFACT 11, XIIIth International Workshop on Neutrino Factories, Super beams and Beta beams, 1-6 August 2011, CERN and University of Geneva
\newline
(Submitted to IOP conference series)}
\end{abstract}

\section{Introduction}
The Booster Neutrino Beamline (BNB) is a conventional neutrino beam at Fermilab which supplies a family of experiments with neutrinos of peak energy around 1GeV.  The MiniBooNE experiment in the BNB has reported an excess of $\nu_e$ like events at low energy in both neutrino and antineutrino mode \cite{AguilarArevalo:2008rc}, and at intermediate energies in antineutrino mode \cite{AguilarArevalo:2010wv}.  Many explanations have been proposed, including new oscillation physics, photon production in neutrino scattering events and the presence of misunderstood instrumental effects.  The MicroBooNE experiment aims to confirm or refute the existence of these excesses with a totally different detector technology.  Unlike MiniBooNE, which cannot differentiate between photon and electron induced Cherenkov rings, the MicroBooNE liquid argon TPC will be able to distinguish between the two event types and hence clarify the nature of the excesses and the evidence they provide for new physics.

The MicroBooNE detector will sit 100m upstream of MiniBooNE in the BNB and plans to accumulate $6.6*10^{20}$ protons on target in neutrino mode.  This will provide enough neutrino interactions to identify the source of the low energy excess, if observed, as photons or electrons with a confidence of $4\sigma$ (photons) or $5 \sigma$ (electrons).  MicroBooNE will also accumulate similar statistics to MiniBooNE in order to search for sterile neutrino oscillations in neutrino mode.   Further, MicroBooNE will perform world-leading measurements of neutrino cross sections on argon.  As well as being of great importance to future liquid argon detectors, these cross section measurements, especially for low track energies, can provide insights into physics topics such as the strange spin of the proton, and the effects of short range nuclear correlations.

The detector is comprised of a stainless steel cryostat filled with 170 tons of liquid argon, of which 70 tons forms the active volume.  This volume is instrumented with a three wireplane time projection chamber (TPC) of 8256 channels and an optical system comprising of 30 photomultiplier tubes assemblies.

In addition to pursuing a diverse program of short baseline neutrino physics, MicroBooNE is an important stepping stone towards a giant liquid argon detector for a future long baseline experiment such as LBNE \cite{Kisslinger:2011fk}.

\section{Operating Principals of Liquid Argon TPCs}
\subsection{Charge Collection}
Neutrino interactions in MicroBooNE produce zero or more charged particles with a momentum in the beam direction.  These particles traverse the active volume and liberate ionization electrons from nearby argon atoms.  Due to the inert nature of argon, the free electron lifetime is long. Under application of a strong enough electric field they can be drifted to one side of the volume.

Along this side are arranged several wireplanes at different angles perpendicular to the drift direction.  MicroBooNE has two induction planes and one collection plane.  Voltages are induced upon the wires by charges passing an induction wire or being absorbed by a collection wire.  By recording of the voltage across each wire, a 3D image of the original charge deposit can be constructed, with the wire numbers providing information about the charge distribution parallel to the wire planes, and the drift time providing the perpendicular coordinate.

\subsection{Light Collection}
Charged particles moving through liquid argon also produce scintillation and Cherenkov light.  MicroBooNE incorporates an optical system to measure this light, which gives several benefits.

The BNB is a pulsed beam, and has structure on a microsecond timescale.  This is also the timescale associated with charge drift through the TPC volume, during which charge losses due to electron absorption, and dispersion via diffusion and space charge effects take place.  In order to obtain the absolute drift distance and apply the corrections necessary for an accurate charge determination, a precise measurement of $t_0$ is required.  The prompt scintillation light gives a $t_0$ measurement with a few ns precision.

An optical trigger also allows for a great simplification in trigger electronics due to the far smaller number of channels present in the optical system, provides stronger rejection of cosmic ray events falling outside the beam window and helps mitigate against TPC noise effects.

An interesting quality of the scintillation light in argon is that it is produced via two pathways, both involving molecular excimer states of argon, one mediated by a singlet excitation and one by a triplet state \cite{Cennini:1999ih}.  These excimers eventually decay into two argon atoms and one 128nm photon, but have very different lifetimes.  As such, there are two time constants associated with the scintillation light: a fast component with $\tau_F = 6ns$ and a slow component with $\tau_S = 1.6 \mu s$.  A competing excimer collision process results in scintillation quenching which is dependent upon the local excimer concentration, and hence the dE/dx of the ionizing particle.  This process more strongly quenches the slow light, hence the ratio of prompt to delayed light is dependent upon the identity and energy of the ionizing particle.  The exploitation of this dependence in augmenting the reconstruction of neutrino events is being investigated.

\section{Current Status of MicroBooNE Subsystems}
\subsection{TPC}
The MicroBooNE TPC system will consist of three wire planes, oriented vertically (3456 wires) and at $\pm 60 \deg$ (2400 wires each).  Each wire is tested for tensile strength and then wound onto a ferrule fixed to a carrier board at each end by an automated wire winding machine.  Wire winding is set to begin this fall at Syracuse and Yale Universities, and boards will be affixed to the TPC frame at Fermilab in Spring 2012.

Simulations of electric field uniformity inside the field cage and the effects of field inhomogeneities upon charge trajectories have been performed.  Image distortions due to field non-uniformity are expected to be no greater than 1mm at any point in the active volume.

\begin{figure}[h]
\begin{minipage}{40pc}
\begin{center}
\includegraphics[height=10pc]{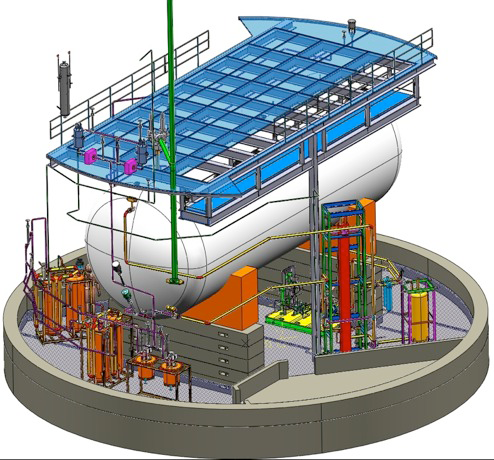}
\includegraphics[height=10pc]{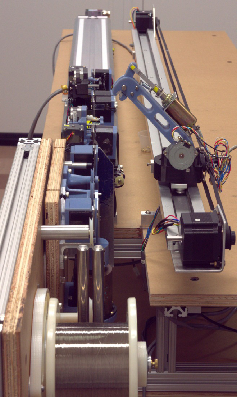}
\includegraphics[height=10pc]{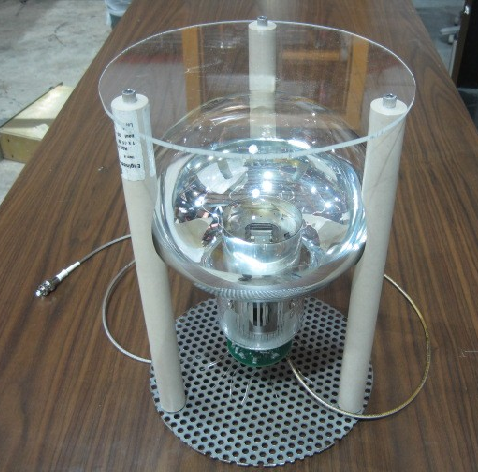}
\includegraphics[height=10pc]{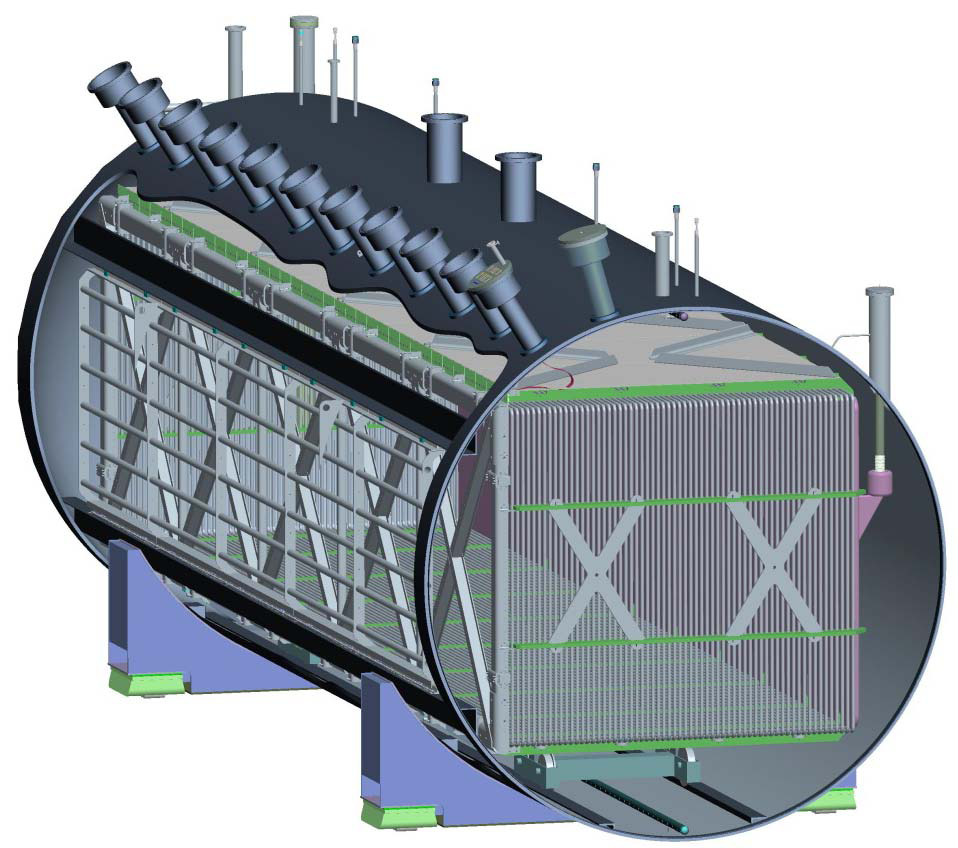}
\end{center}
\caption{\label{label}Left to Right: The MicroBooNE cryostat in situ with surrounding cryogenics,  the BNL automated wire winding machine, a mechanical model of one optical system PMT assembly, cutaway of cryostat showing field cage and feedthrough arrangement} 
\end{minipage}
\end{figure}

\subsection{Light Collection System}
The MicroBooNE optical system is comprised of 30 8 inch cryogenic photomultiplier tubes (PMTs) mounted behind acrylic plates which are coated with a layer of tetraphenyl butadiene (TPB) enriched polystyrene.  TPB is a wavelength shifter which absorbs light over a wide range of wavelengths and emits in the visible.  One PMT, wavelength shifting plate, plastic support stand, cryogenic base and combined signal and high voltage cable comprises one self contained PMT assembly.  

The PMT assemblies will be arranged on a rack which slides behind the TPC wireplanes, a position chosen to avoid exposing the PMTs to high field environments.  Testing of each of the MicroBooNE PMTs in cryogenic conditions has begun at Fermilab in a liquid nitrogen test stand, and a prolonged vertical slice test of several full assemblies will be performed in liquid argon in the new year in order to test the stability and performance of the units.

\subsection{Cryogenics}
MicroBooNE employs an extensive system of cryogenics in order to maintain argon purity.  Oxygen and water continuously outgas from the cryostat, and trace amounts can greatly reduce free electron lifetime. Concentration of these impurities must be held below 100ppt. To achieve this, constant purification is performed, and the electron lifetime is measured in real time by lifetime monitors both inside the cryostat and in the recirculation system.  Scintillation light is also quenched by nitrogen, and the dissolved nitrogen level must be kept below 100ppb.  

In order to prevent image distortions due to nonuniform drift velocities, temperature uniformity must be maintained to within 0.1K across the fiducial volume.  Computational fluid dynamical simulations have been performed to model the convection currents and heat flow in the detector and confirm that such a uniformity can be readily achieved, with convection flow rates being negligible compared to the electron drift velocity.

\subsection{Electronics}
MicroBooNE will employ both warm and cold electronics, in a scheme designed to minimize TPC noise.  Initial preamplification and signal shaping is performed by cold front end ASICs inside the cryostat.  The signal is transported to several cryostat feedthroughs by a twisted pair, at which point an intermediate level of amplification is performed by warm interface electronics.  Cables from the interface run 20-30m to a DAQ system in the detector hall.

\subsection{Computing and Software}

Simulation, reconstruction and data analysis jobs for MicroBooNE are performed using the LArSoft software package, which is a general purpose framework for liquid argon TPC experiments  \cite{LArSoft}.  LArSoft is developed and maintained by collaborators from the participating experiments, currently ArgoNeut \cite{ArgoNeuT}, MicroBooNE, LBNE LAr and some Fermilab test stands.

LArSoft interfaces with external event generators to simulate neutrino beam interactions \cite{Andreopoulos:2009rq} and cosmic ray events \cite{CRY}, and also includes internal event generators for validation studies.  There is a full GEANT4 simulation, customized to provide fast handling of charge drift physics and voltage induction in the TPC wires.  LArSoft also incorporates optical simulation capabilities, with both full photon tracking and fast volume sampling based simulation methods.

The vast amount of information recorded for a single neutrino interaction event in liquid argon makes the task of achieving an optimal event reconstruction formidable.  The development of new algorithms by the LArSoft and MicroBooNE collaborations is ongoing, and whilst basic 3D tracking and shower finding methods are in place, work on reconstruction methods of increasing sophistication will continue into the analysis phase of the experiment.

\begin{figure}[h]
\begin{minipage}{40pc}
\begin{center}
\includegraphics[width=6.5pc]{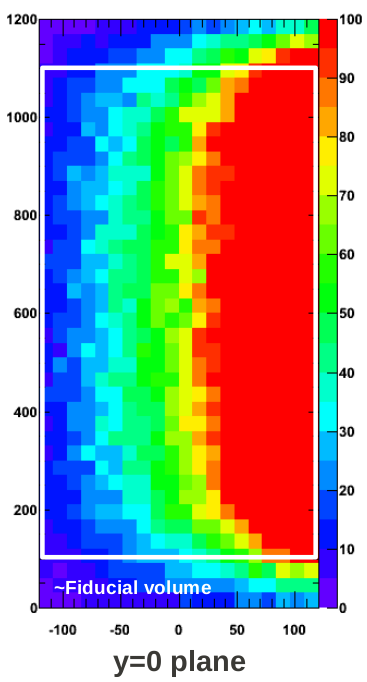}
\includegraphics[width=6.5pc]{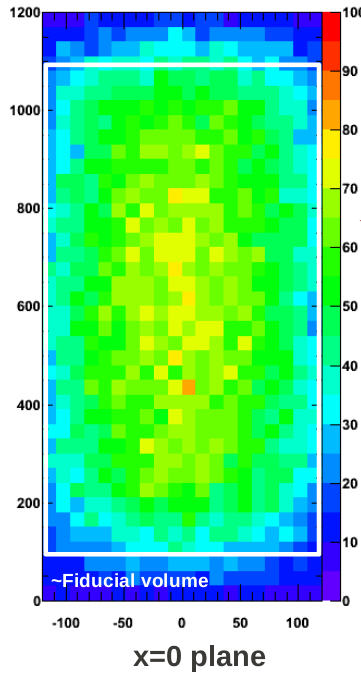}
\hspace{2pc}
\includegraphics[width=14pc]{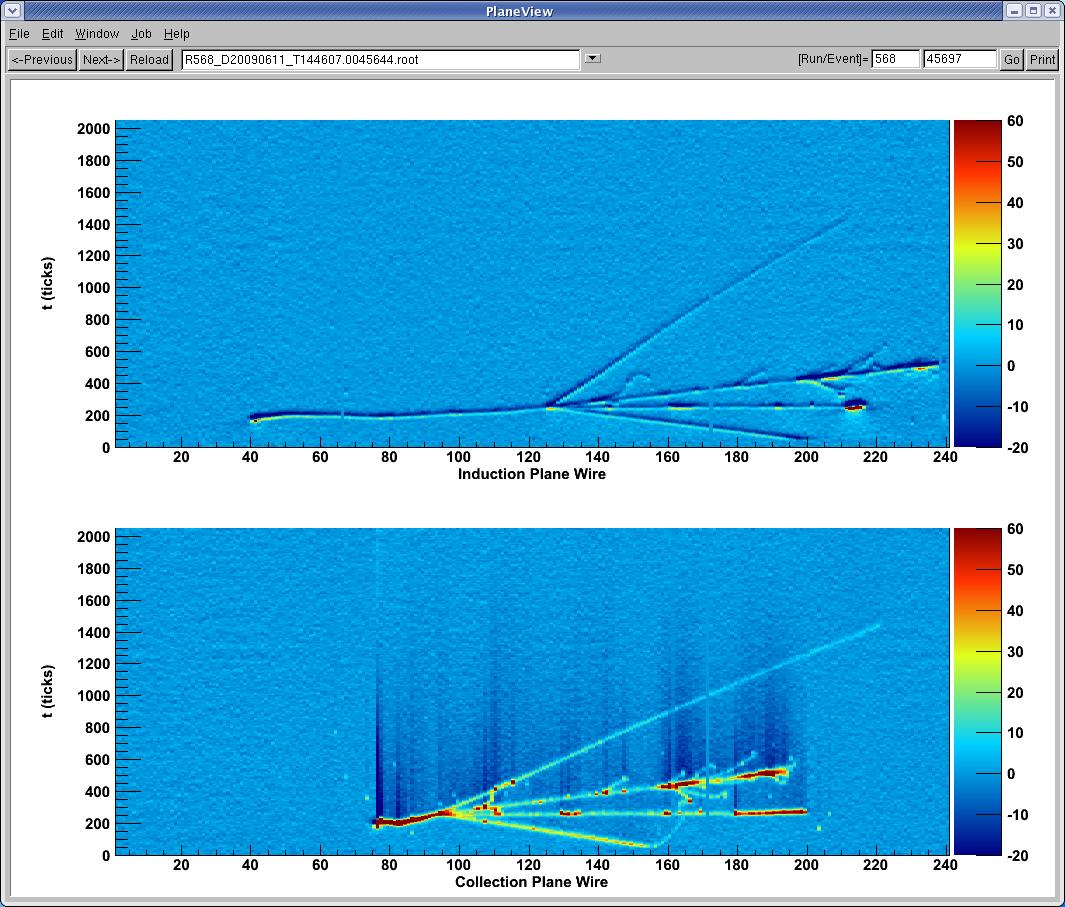}
\end{center}
\caption{\label{label}Left : Trigger sensitivity map for the x=0 and y=0 detector planes, produced using the LArSoft optical physics simulation.  Right : Event display showing raw data for a typical neutrino event from the ArgoNeuT experiment} 
\end{minipage}
\end{figure}

\section{Summary}
The MicroBooNE experiment is a 170T liquid argon neutrino experiment which is set to begin running in the BNB at Fermilab in early 2014.  Construction and testing of both the TPC and optical detector systems are underway, and engineering designs and prototypes for cryogenic systems and electronics are in place.  The experiment will pursue a range of short baseline physics including cross section measurements and an exploration of the MiniBooNE anomalies with an emerging and powerful neutrino detection technology.

\section*{References}

\bibliography{bib-nufact}

\end{document}